\documentclass[pdflatex,sn-mathphys-num]{sn-jnl}% Math and Physical Sciences 
\usepackage{graphicx}%
\usepackage{multirow}%
\usepackage{amsmath,amssymb,amsfonts}%
\usepackage{amsthm}%
\usepackage{mathrsfs}%
\usepackage[title]{appendix}%
\usepackage{xcolor}%
\usepackage{textcomp}%
\usepackage{manyfoot}%
\usepackage{booktabs}%
\usepackage{algorithm}%
\usepackage{algorithmicx}%
\usepackage{algpseudocode}%
\usepackage{listings}%
\theoremstyle{thmstyleone}%
\theoremstyle{thmstyletwo}%

\theoremstyle{thmstylethree}%

\begin{document}

\title[Bayesian perspectives on exponential random graph models]{Bayesian perspectives on exponential random graph models}

\author*[]{\fnm{Alberto} \sur{Caimo}}\email{alberto.caimo1@ucd.ie}

\author[ ]{\fnm{Isabella} \sur{Gollini}}

\affil[ ]{\orgdiv{School of Mathematics and Statistics},
          \orgname{University College Dublin},
          \country{Ireland}}

\abstract{Exponential random graph models (ERGMs) are a widely used framework for network data, enabling hypothesis testing on the structural mechanisms underlying observed networks. Bayesian ERGMs provide principled uncertainty quantification and enable the incorporation of prior knowledge through fully probabilistic modelling. However, computation remains challenging because the posterior is doubly intractable, with a likelihood normalising constant that depends on unknown parameters.
This paper reviews Bayesian approaches to ERGM inference, categorising inference methods into three broad classes: auxiliary variable MCMC methods, adjusted pseudo-likelihood approaches, and variational methods, alongside dedicated treatment of model selection. We also discuss modelling extensions for missing data, longitudinal dynamics, populations of networks, weighted networks, highlighting applications across various scientific disciplines.}
\keywords{Exponential random graphs, Bayesian analysis, approximate inference, doubly-intractable distributions}

\maketitle

\section{Introduction}\label{introduction}

Exponential random graph models (ERGMs, \citep{fra:str86}) are statistical models for networks that treat the observed network graph as a random variable, enabling inference on network structure. They are widely used, particularly in sociology, epidemiology, political science, and organisational research. In many applications, researchers have prior hypotheses about mechanisms such as homophily, reciprocity, and transitivity. Bayesian formulations allow these beliefs to be encoded through priors while providing principled uncertainty quantification. Early work, such as that of \citet{kos04} and \citet{cai:fri11}, established the theoretical foundations for a Bayesian treatment of ERGMs and showed that Bayesian frameworks can characterise ERGMs rigorously providing regularisation and full uncertainty quantification.

ERGMs characterise the probability distribution of network graphs by modelling the topological structure of an observed network $\mathbf{y} \in \mathcal{Y}$ on $n$ nodes, represented by its $n \times n$ adjacency matrix, through a set of summary network statistics $s(\mathbf{y}) \in \mathbb{R}^p.$ These statistics can flexibly incorporate a wide range of structural features as well as nodal and/or dyadic covariates. 
The ERGM likelihood is given by
$$p(\mathbf{y} \mid \boldsymbol{\vartheta}) = 
\frac{\exp\left\{\boldsymbol{\vartheta}^\top s(\mathbf{y})\right\}}
     {\kappa(\boldsymbol{\vartheta})},$$
where $\boldsymbol{\vartheta} \in \mathbb{R}^p$ is a vector of parameters associated with the statistics, and $\kappa(\boldsymbol{\vartheta})$ is the normalising constant of the form,
$$\kappa(\boldsymbol{\vartheta}) = \sum_{\mathbf{y} \in \mathcal{Y}} \exp\left\{\boldsymbol{\vartheta}^\top s(\mathbf{y})\right\},$$
which is typically intractable due to the combinatorially large space $\mathcal{Y}$ of all possible network configurations on $n$ nodes; this, in turn, makes the sampling distribution analytically intractable and constitutes the main difficulty in fitting ERGMs.

In the Bayesian setup, we define a prior distribution on the parameters $\pi(\boldsymbol{\vartheta})$ and we target the posterior distribution,
$$
\pi(\boldsymbol{\vartheta} \mid \mathbf{y}) = \frac{p(\mathbf{y} \mid \boldsymbol{\vartheta})\;\pi(\boldsymbol{\vartheta})}{p(\mathbf{y})},
$$
which is doubly intractable due to the additional intractability of the model evidence, or marginal likelihood, $p(\mathbf{y}).$

This review paper will survey some of the main advances in the Bayesian analysis of ERGMs by exploring computational approaches for parameter and model inference (Section~\ref{computational_methods}) and modelling approaches for complex heterogenous network data (Section~\ref{modelling}). We will also review the main software tools available (Section~\ref{software}). We will conclude this review with a discussion and future directions (Section~\ref{discussion}), covering computational scalability, modelling extensions, and open challenges.

\section{Computational Methods}\label{computational_methods}

\subsection{Parameter Inference}

The problem of estimating doubly-intractable posterior distributions, such as the ERGM posterior $\pi(\boldsymbol{\vartheta} \mid \mathbf{y})$, has become a major topic in computational statistics. A comprehensive review of Bayesian methods for handling intractable normalising constants is provided by \citet{par:har18}. The following sections summarise the principal approaches used in Bayesian ERGM parameter inference: auxiliary variable methods, which introduce latent variables to cancel the normalising constants in the MCMC acceptance ratio, and likelihood approximation methods, which replace $\kappa(\boldsymbol{\vartheta})$ with a tractable surrogate, including pseudo-likelihood and variational approaches.

\subsubsection{Auxiliary Variable MCMC}

In the Bayesian setup, a common ERGM estimation procedure is the approximate exchange algorithm introduced by \citet{cai:fri11} by adapting the algorithm from \citet{Murray06} approximating the ERGM likelihood with 
at each step of a MCMC algorithm proposing a new parameter $\boldsymbol{\vartheta}' \sim h(\cdot \mid \boldsymbol{\vartheta})$ (e.g., symmetric proposal) and subsequent MCMC draw of simulated network data $\mathbf{y}' \sim p(\cdot \mid \boldsymbol{\vartheta}')$ and targeting an augmented joint distribution,
$\pi(\boldsymbol{\vartheta}, \boldsymbol{\vartheta}', \mathbf{y}, \mathbf{y}') \propto \pi(\boldsymbol{\vartheta}' \mid \mathbf{y}')\; \pi(\boldsymbol{\vartheta} \mid \mathbf{y}).$
The key computational advantage of this construction is that the intractable normalising constants $\kappa(\boldsymbol{\vartheta})$ and $\kappa(\boldsymbol{\vartheta}')$ cancel in the Metropolis--Hastings acceptance ratio, which takes the form
$$\min\left(1,\; \frac{\pi(\boldsymbol{\vartheta}')}{\pi(\boldsymbol{\vartheta})} \cdot \exp\Big\{(\boldsymbol{\vartheta}' - \boldsymbol{\vartheta})^\top \big[s(\mathbf{y}) - s(\mathbf{y}')\big]\Big\}\right).$$
Since exact simulation from $\mathbf{y}' \sim p(\cdot \mid \boldsymbol{\vartheta}')$ is itself
infeasible, \citet{cai:fri11} replace it with an approximation obtained by
running an inner MCMC chain, typically a Gibbs sampler over edge toggles,
initialised at the observed network $\mathbf{y}$, rendering the procedure
approximate but computationally feasible \citep{eve12}. 

Further efficiency improvements have been proposed by \citet{cai:mir15} using a synthesis of delayed rejection and adaptive Monte Carlo methods and by \citet{alq:fri:eve:bol16} using noisy Monte Carlo with approximate transition kernels which they proved still converge to a neighbourhood of the correct posterior under controlled noise conditions. 
In order to further improving asymptotic exactness, \citet{liaetal16} proposed an adaptive exchange algorithm which uses an importance sampling approximation: a pool of $m$ auxiliary pairs $\{(\boldsymbol{\vartheta}_j, \mathbf{y}_j)\}_{j=1}^m$ is maintained from a parallel chain, and the intractable ratio is approximated as
$$\frac{\kappa(\boldsymbol{\vartheta})}{\kappa(\boldsymbol{\vartheta}')} \approx \sum_{j=1}^m w_j \exp\!\Big[(\boldsymbol{\vartheta} - \boldsymbol{\vartheta}')^\top s(\mathbf{y}_j)\Big],$$
with importance weights $w_j \propto \exp\left(\boldsymbol{\vartheta}_j^\top s(\mathbf{y}_j)\right)$ where the pool is updated adaptively as sampling proceeds.

\citet{lia:jin13} proposed a Monte Carlo Metropolis-Hastings algorithm that approximates the intractable normalising constant ratio $\kappa(\boldsymbol{\vartheta}')/\kappa(\boldsymbol{\vartheta})$ via importance sampling from a short inner Markov chain, substituting this estimate directly into the acceptance ratio. Unlike exchange-type algorithms, perfect sampling from the model distribution is not required, and convergence to the correct posterior is established under mild regularity conditions.

A recent and significant advance was proposed by \citet{zha:lia24}, who adapted Stochastic Gradient Langevin Dynamics (SGLD) by \citet{wel:teh11} to the doubly-intractable ERGM posterior, yielding a method that retains theoretical convergence guarantees.
All exact Bayesian ERGM methods require an inner MCMC chain to simulate an auxiliary network $\mathbf{y}' \sim p(\cdot \mid \boldsymbol{\vartheta}')$ at every iteration. \citet{zha:lia24} sidestep this requirement by abandoning the accept-reject correction entirely and instead exploiting the Langevin gradient structure. The SGLD update takes the form
$$\boldsymbol{\vartheta}_{t+1} = \boldsymbol{\vartheta}_t + \frac{\epsilon_t}{2}\mathbf{D}\nabla_{\boldsymbol{\vartheta}} \log \pi(\boldsymbol{\vartheta}_t \mid \mathbf{y}) + \boldsymbol{\eta}_t, \qquad \boldsymbol{\eta}_t \sim \mathcal{N}(\mathbf{0}, \epsilon_t\, \mathbf{D}),$$
where $\epsilon_t$ is a positive scalar and $\mathbf{D}$ is a $d \times d$ diagonal matrix,  which together give component-specific learning rates for $\boldsymbol{\vartheta},$ and the gradient of the log-posterior decomposes as $\nabla_{\boldsymbol{\vartheta}} \log \pi(\boldsymbol{\vartheta} \mid \mathbf{y}) = s(\mathbf{y}) - \mathbb{E}_{\boldsymbol{\vartheta}}[s(\mathbf{Y})] + \nabla_{\boldsymbol{\vartheta}} \log \pi(\boldsymbol{\vartheta}).$ At each iteration, $\mathbb{E}_{\boldsymbol{\vartheta}}[s(\mathbf{Y})]$ is approximated using a short inner Markov chain of length $m$, avoiding explicit evaluation of the intractable normalising constant $\kappa(\boldsymbol{\vartheta})$. Since inference proceeds through stochastic gradient updates rather than a Metropolis--Hastings acceptance step, exact auxiliary network simulation is not required. The method admits non-asymptotic convergence guarantees in 2-Wasserstein distance for any fixed $m \geq 1$, whereas exchange-based algorithms recover exactness only as the inner chain length increases. Empirically, SGLD achieves accuracy comparable to the adaptive exchange algorithm of \citet{liaetal16} at substantially lower computational cost, while outperforming double Metropolis--Hastings sampling.

\subsubsection{Adjusted Pseudo-likelihood}

The pseudo-likelihood function factorises the joint ERGM distribution as a product of conditional Bernoulli terms over each dyad $(i,j)$:
$$p_{PL}(\boldsymbol{\vartheta} \mid \mathbf{y}) = \prod_{i<j} p(y_{ij} \mid \mathbf{y}_{-ij}, \boldsymbol{\vartheta}) = \prod_{i<j} \frac{\exp\left(\boldsymbol{\vartheta}^\top \delta_{ij}(\mathbf{y})\, y_{ij}\right)}{1 + \exp\left(\boldsymbol{\vartheta}^\top \delta_{ij}(\mathbf{y})\right)}$$
where $\delta_{ij}(\mathbf{y}) = s(\mathbf{y}^+_{ij}) - s(\mathbf{y}^-_{ij})$ is the change statistic for toggling the value of dyad $(i,j)$. This makes the normalising constant $\kappa(\boldsymbol{\vartheta})$ tractable, but at the cost of assuming dyadic independence, a serious misspecification whenever the model includes transitivity or other dependence terms.

\citet{bou:fri:mai17} perform three adjustments to the pseudo-likelihood: (1) mode correction, to overcome the bias of the pseudo-posterior mode; (2) curvature adjustment, modifying the transformation matrix and the corresponding Hessian; and (3) magnitude adjustment, a linear rescaling to bring the curvature-adjusted pseudo-likelihood to the correct scale.
The adjusted log-pseudo-likelihood applies an affine correction,
$$
\log \tilde{p}_{PL}(\boldsymbol{\vartheta} \mid \mathbf{y}) 
= \log p_{PL}(\mathbf{C}\boldsymbol{\vartheta} + d \mid \mathbf{y}),
$$
where $d$ recentres the mode from $\hat{\boldsymbol{\vartheta}}_{\mathrm{MPLE}}$ to $\hat{\boldsymbol{\vartheta}}_{\mathrm{MLE}}$, and $\mathbf{C}$ rescales curvature using the ratio of Hessians at the respective modes. The MLE is obtained via the standard MCMC-MLE scheme of \citet{gey:tho92}. The resulting adjusted pseudo-posterior,
$\tilde{\pi}(\boldsymbol{\vartheta} \mid \mathbf{y}) \propto 
\tilde{p}_{PL}(\boldsymbol{\vartheta} \mid \mathbf{y}) \, \pi(\boldsymbol{\vartheta}),$
can be explored with standard Metropolis--Hastings using only pseudo-likelihood evaluations, yielding a computationally efficient approximation.

\subsubsection{Variational Methods}

\citet{tan:fri20} developed a variational Bayesian framework for ERGMs that replaces MCMC-based posterior sampling with an optimisation problem, targeting a tractable approximation to the doubly intractable posterior at substantially reduced inferential computational cost. The core idea is to approximate the true posterior $\pi(\boldsymbol{\vartheta} \mid \mathbf{y})$ by a Gaussian variational distribution $q(\boldsymbol{\vartheta}) = \mathcal{N}(\boldsymbol{\mu}, \boldsymbol{\Sigma})$, found by minimising the Kullback–Leibler divergence, $\mathrm{KL}(q\ \|\ \pi),$ which is equivalent to maximising the evidence lower bound (ELBO). Since $\log \pi(\boldsymbol{\vartheta} \mid \mathbf{y})$ involves $\kappa(\boldsymbol{\vartheta})$, which is itself intractable, direct evaluation of the ELBO is infeasible. To overcome the computational hurdle of drawing a network from the likelihood at each iteration, \citet{tan:fri20} propose stochastic gradient ascent with consistent gradient estimates computed using adaptive self-normalised importance sampling. Two algorithms are developed: (i) non-conjugate variational message passing (NCVMP) based on the adjusted pseudo-likelihood of Bouranis et al. (2017), where the intractable likelihood is replaced by the calibrated $\tilde{p}_{PL}(\boldsymbol{\vartheta} \mid \mathbf{y})$ before computing the ELBO gradient, making each iteration cheap since no inner simulation is required; and (ii) stochastic variational inference (SVI), which retains the true likelihood but approximates its gradient at each step via a small batch of networks simulated from the ERGM likelihood via MCMC, yielding biased but consistent gradient estimates.
The NCVMP update equations are derived by fixing $q(\boldsymbol{\vartheta}_{-k})$ and optimising over each component $\boldsymbol{\vartheta}_k$ in turn, giving closed-form natural gradient steps for the Gaussian variational parameters $(\boldsymbol{\mu}, \boldsymbol{\Sigma})$. \citet{tan:fri20} developed a variety of variational methods for Gaussian approximation of the posterior density and model selection, shown to yield
comparable performance to that of the approximate exchange algorithm.

\subsection{Model Selection}

Model selection in ERGMs concerns the choice of which sufficient statistics $s(\mathbf{y})$ to include in the model, equivalently, which network features are genuinely supported by the observed network.

This represents a substantive scientific question: for example, including a transitivity term implies that clustering acts as a generative mechanism in the network beyond what would be expected from density alone. 
For ERGMs, the obstacle is the same intractable normalising constant $\kappa(\boldsymbol{\vartheta})$ that complicates parameter estimation: the model evidence or marginal likelihood, which represents the probability of the observed data under model $\mathcal{M}_k,$ is given by
$$p(\mathbf{y} \mid \mathcal{M}_k) = \int \frac{\exp\left(\boldsymbol{\vartheta}^\top s_k(\mathbf{y})\right)}{\kappa_k(\boldsymbol{\vartheta})} \, \pi(\boldsymbol{\vartheta}) \, \mathrm{d}\boldsymbol{\vartheta}.$$
Evaluating this quantity involves both integration over the parameter space and the intractable normalising constant $\kappa_k(\boldsymbol{\vartheta})$ inside the integrand, a doubly intractable quantity that is model-specific and cannot be cancelled. 

Bayesian model selection proceeds via the Bayes factor (BF) between competing models $\mathcal{M}_k$ and $\mathcal{M}_{k'}$:
$$\mathrm{BF}_{kk'} = \frac{p(\mathbf{y} \mid \mathcal{M}_k)}{p(\mathbf{y} \mid \mathcal{M}_{k'})}$$
which directly quantifies the relative evidence the observed network provides for each model, averaging over parameter uncertainty rather than conditioning on a point estimate. Combined with prior model probabilities $p(\mathcal{M}_k)$, it yields posterior model probabilities and enables formal model averaging.

\subsubsection{Auxiliary Variable Reversible Jump MCMC}

\citet{cai:fri13} extended the approximate exchange algorithm to a reversible jump (RJ) MCMC framework, enabling simultaneous inference over both the parameter space and the model space. The fundamental obstacle is that standard RJ-MCMC   \citep{gre95} requires evaluation of the likelihood ratio between competing models, which for ERGMs involves the ratio of two model-specific normalising constants $\kappa_k(\boldsymbol{\vartheta}_k)$ and $\kappa_{k'}(\boldsymbol{\vartheta}_{k'})$, quantities that are individually intractable and, crucially, do not cancel across models of different dimension as they do in the within-model exchange algorithm.

The proposed sampler explores the augmented space $(\mathcal{M}_k, \boldsymbol{\vartheta}_k)$ via two types of move. Within-model moves proceed via the standard approximate exchange update of \cite{cai:fri11}. Between-model moves propose a jump to a competing model $\mathcal{M}_{k'}$ of different dimension following \citet{gre95}: auxiliary variables $\mathbf{u}$ are drawn from a proposal density $q(\mathbf{u})$ to match dimensions, and an auxiliary network $\mathbf{y}' \sim p(\cdot \mid \boldsymbol{\vartheta}_{k'}, \mathcal{M}_{k'})$ is simulated under the proposed model via an inner MCMC chain. 
The empirical frequencies of model visits can be used to estimate the posterior model probabilities $p(\mathcal{M}_k \mid \mathbf{y})$ for each competing model.

\subsubsection{Adjusted Pseudo-likelihood}

The adjusted pseudo-likelihood approach proposed by \citet{bou:fri:mai17} can be used to estimate the model evidence.  \citet{bou:fri:mai18} extended their parameter inference framework by using thermodynamic integration over a power posterior ladder $\{\tilde{\pi}^t(\boldsymbol{\vartheta} \mid \mathbf{y})\}_{t \in [0,1]}$, enabling Bayes factor comparisons between competing models. This approach turned out to be  particularly valuable since computing evidence under the true intractable likelihood is essentially infeasible.

\subsubsection{Variational Methods}

The variational approach of \citet{tan:fri20} can be used for model comparison. In this case, the key quantity is the log model evidence $\log p(\mathbf{y} \mid \mathcal{M}_k)$, which is lower bounded by the ELBO:
$$\log p(\mathbf{y} \mid \mathcal{M}_k) \geq \mathcal{L}(q) =
\mathbb{E}_q[\log \tilde{p}_{PL}(\boldsymbol{\vartheta} \mid \mathbf{y})]
+ \mathbb{E}_q[\log \pi(\boldsymbol{\vartheta})] - \mathbb{E}_q[\log
q(\boldsymbol{\vartheta})].$$
Here the ELBO is expressed using the adjusted pseudo-likelihood $\tilde{p}_{PL}$
as in the NCVMP variant of \citet{tan:fri20}; the SVI variant constructs an
analogous bound in which the true ERGM likelihood gradient is approximated
via inner MCMC simulation, at greater computational cost but without the
pseudo-likelihood approximation error.
Since $q$ is Gaussian and the adjusted pseudo-likelihood is available in
closed form, all three expectations are either available analytically or
cheaply approximated via Monte Carlo quadrature. The ELBO thus serves as a
tractable surrogate for $\log p(\mathbf{y} \mid \mathcal{M}_k)$, and Bayes
factors between competing models are approximated as
$\log \widehat{\mathrm{BF}}_{kk'} \approx \mathcal{L}_k(q) - \mathcal{L}_{k'}(q).$
This entirely avoids thermodynamic integration, power posteriors, or the
reversible jump construction, reducing model selection to independent
variational optimisation runs for each candidate model.

\section{Modelling Extensions}\label{modelling}

\subsection{Missing Data}

\citet{kos:rob:pat10} addressed the problem of missing data in ERGMs, which is particularly consequential in network analysis since unobserved ties alter local neighbourhood structures and thereby distort inference on all model parameters. Under the assumption of data missing at random, the observed-data likelihood is obtained by marginalising the complete-data ERGM likelihood over the unobserved ties, a strategy adapted from the general missing data framework of \citet{rub76}. Inference proceeds via Bayesian data augmentation: the missing tie variables $\mathbf{y}_{\text{mis}}$ are treated as latent quantities and sampled jointly with the model parameters $\boldsymbol{\vartheta}$ within a Gibbs scheme, with the full conditional of $\mathbf{y}_{\text{mis}}$ given $\boldsymbol{\vartheta}$ available as a standard ERGM distribution on the missing subgraph. This simultaneously yields posterior inference on the parameters and predictive distributions over the unobserved ties, enabling principled network imputation. The framework was subsequently extended by \citet{kos:rob:wan:pat13} to accommodate missing node attributes and partially observed actor sets.

\subsection{Nodal Heterogeneity}

\citet{thi:fri:cai:kau16} extend the Bayesian ERGM framework of \citet{cai:fri11} to account for unobserved heterogeneity across nodes, a limitation of standard ERGMs that assume nodal homogeneity, an assumption frequently violated in practice when networks exhibit degree heterogeneity or scale-free behaviour. The proposed model augments the standard ERGM with node-specific random effects $\boldsymbol{\alpha} = (\alpha_1, \ldots, \alpha_n)$, yielding a mixed ERGM of the form $p(\mathbf{y} \mid \boldsymbol{\vartheta}, \boldsymbol{\alpha}) \propto \exp\!\left(\boldsymbol{\vartheta}^\top s(\mathbf{y}) + \boldsymbol{\alpha}^\top d(\mathbf{y})\right)$ where $d(\mathbf{y})$ collects the node degrees and the random effects $\alpha_i \sim \mathcal{N}(0, \sigma^2)$ are assigned a Gaussian prior, drawing on the $p_2$ model of \citet{van:sni:zij04}. The full posterior over $(\boldsymbol{\vartheta}, \boldsymbol{\alpha}, \sigma^2)$ is then doubly intractable in the same sense as the standard ERGM posterior, since the normalising constant $\kappa(\boldsymbol{\vartheta}, \boldsymbol{\alpha})$ is intractable. Inference proceeds via an extension of the approximate exchange algorithm, where the auxiliary network $\mathbf{y}'$ is simulated conditionally on both the proposed structural parameters $\boldsymbol{\vartheta}'$ and the random effects $\boldsymbol{\alpha}$, and the random effects are updated via a standard Gibbs step given the current network and structural parameters. For model selection, the paper develops a Laplace approximation to the marginal likelihood, integrating out $\boldsymbol{\alpha}$ analytically under the Gaussian prior and applying path sampling to approximate the remaining intractable normalising constant. This enables Bayes factor comparisons between the pure $p_2$ model, the pure ERGM, and the combined mixed ERGM.

A complementary approach to modelling unobserved heterogeneity was proposed by \citet{box:ste:chr:mor18}, who introduced the Frailty Exponential Random Graph Model (FERGM). Rather than the Gaussian nodal random effects of \citet{thi:fri:cai:kau16}, the FERGM incorporates node-specific frailty terms as additive components in the linear predictor, capturing unexplained variation in the propensity of nodes to send and receive edges. 

\subsection{Local Dependence}

\citet{sch:han15} address a fundamental theoretical limitation of standard ERGMs: 
conventional dependence assumptions, such as those of Markov random graphs, induce 
\textit{global} dependence, rendering such models near-degenerate and statistically 
ill-behaved on large networks. ERGMs with \textit{local 
dependence} assume that the node set is partitioned into $K$ non-overlapping 
neighbourhoods, with dependence permitted only within neighbourhoods and edges across 
neighbourhoods remaining mutually independent. The log-normalising constant 
consequently factorises as
$$\kappa(\boldsymbol{\vartheta}, \mathbf{z}) 
= \sum_{k=1}^{K} \kappa_{W,k}(\boldsymbol{\vartheta}_{W,k}) 
+ \sum_{k=1}^{K}\sum_{l<k}\sum_{i \in A_k,\, j \in A_l} 
\kappa_{B,i,j}(\boldsymbol{\vartheta}_B),$$
where $\kappa_{W,k}$ and $\kappa_{B,i,j}$ are the within- and between-neighbourhood 
log-normalising constants, respectively. When neighbourhood structure is unobserved, 
\citet{sch:han15} adopt a Bayesian approach: a truncated stick-breaking prior is placed 
on membership probabilities, Gaussian priors on the parameter vectors, and conjugate 
hyperpriors on the mean and precision of the Gaussian base distribution. Since the 
within-neighbourhood constants $\kappa_{W,k}$ are intractable, posterior inference 
proceeds via the approximate exchange algorithm of \citet{cai:fri11}. The 
block-wise factorisation of $\kappa(\cdot)$ is directly exploited here, as auxiliary graphs 
need only be simulated block by block rather than over the full network, substantially 
reducing the computational cost relative to globally dependent ERGMs.

\subsection{Multiple Networks}

In many applications, researchers observe not a single network but a collection of related networks. Such settings require methods that can jointly model shared structural tendencies while accounting for heterogeneity between networks.
\citet{sla:koe16} developed one of the first fully Bayesian hierarchical ERGM frameworks for samples of networks, enabling inference on both network-specific structural tendencies and variation across groups. Individual networks are modelled through standard ERGMs, while network-specific parameters are assumed to arise from a common population distribution, inducing partial pooling across networks and yielding more stable estimates than fitting separate ERGMs independently. The framework also accommodates network-level covariates to assess how contextual factors moderate structural effects. Inference is performed by embedding the approximate exchange algorithm of \citet{cai:fri11} within a Gibbs sampler, with individual-level parameters updated via exchange steps and population-level hyperparameters sampled conditionally.

The work was subsequently generalised by \citet{leh:etal21} and \citet{leh:whi24}, who developed a Bayesian hierarchical framework that extends the ERGM from the single-network setting to the analysis of \textit{populations of networks}, motivated by applications in neuroimaging where one observes $N$ networks $\{\mathbf{y}^{(i)}\}_{i=1}^N,$ one per individual, rather than a single graph. Each data point can be seen as a realisation of a network-valued random variable, and each may be accompanied by covariate information whose effect on network structure one wishes to assess across the population.
The hierarchical model assigns to each individual network $\mathbf{y}^{(i)}$
its own ERGM parameter $\boldsymbol{\vartheta}^{(i)}$, which is itself drawn
from a population-level distribution governed by shared hyperparameters
$\boldsymbol{\phi} = (\boldsymbol{\mu}, \boldsymbol{\Sigma})$:
$$p(\mathbf{y}^{(i)} \mid \boldsymbol{\vartheta}^{(i)}) \propto
\exp\!\left(\boldsymbol{\vartheta}^{(i)\top} \mathbf{s}(\mathbf{y}^{(i)})
\right), \qquad \boldsymbol{\vartheta}^{(i)} \sim \mathcal{N}(\boldsymbol
{\mu}, \boldsymbol{\Sigma}),$$
with hyperpriors placed on $\boldsymbol{\phi}$. This structure pools
information across networks, providing more precise individual-level estimates,
whilst accommodating genuine between-subject variability in network topology.
To perform inference for populations of networks, the authors develop an exchange-within-Gibbs algorithm combining the exchange algorithm with Gibbs sampling and an ancillarity-sufficiency interweaving strategy to improve mixing. Individual-level parameters $\boldsymbol{\vartheta}^{(i)}$ are updated via exchange steps requiring auxiliary network simulations, while population-level parameters $\boldsymbol{\phi}$ are sampled conditionally via Gibbs updates. Since the $\boldsymbol{\vartheta}^{(i)}$ are conditionally independent given $\boldsymbol{\phi}$, updates can be parallelised, mitigating the linear increase in computational cost with the number of individuals $N$. The method is illustrated on resting-state fMRI data from the Cam-CAN ageing study to compare functional brain connectivity across age groups, extending Bayesian ERGM inference to replicated network settings.

Whereas the hierarchical model of \citet{leh:whi24} captures between-subject variation through a common population distribution, \citet{yin:she:but22} instead model \textit{ensembles of networks} using a finite mixture of ERGMs, allowing for heterogeneous generative mechanisms across graphs. Each mixture component is an ERGM with parameters $\boldsymbol{\vartheta}_k$, while each observed network $\mathbf{y}^{(i)}$ is assigned a latent cluster label $Z^{(i)}$ with mixing weights $\boldsymbol{\tau}$. To address the intractability of the ERGM likelihood, the authors employ pseudo-likelihood and adjusted pseudo-likelihood approximations, with inference performed via a Metropolis-within-Gibbs sampler over cluster assignments, component parameters, and mixing weights. The number of components is estimated through deliberate overfitting with sparse Dirichlet priors, shrinking redundant clusters towards zero. The approach supports model-based clustering and density estimation for heterogeneous network ensembles and is illustrated on simulated and U.S. Senate co-voting networks.

\citet{ren:etal23} extended the finite ERGM mixture framework of \citet{yin:she:but22} to a Bayesian nonparametric setting via a Dirichlet Process Mixture of ERGMs, allowing the number of network clusters to be determined automatically from the data. Inference employs an intermediate importance sampling technique inside a Metropolis-within-slice sampling scheme, addressing the problem of sampling from intractable ERGMs on an infinite sample space. Three likelihood strategies, exact, pseudo-likelihood, and adjusted pseudo-likelihood, are considered, offering a practical trade-off between accuracy and computational cost.

\citet{yin:but22} developed a highly scalable framework for maximum likelihood and conjugate Bayesian inference for ERGMs on \textit{graph sets with equivalent vertices}, i.e., collections of networks with exchangeable nodes. Under this assumption, the pooled sufficient statistic reduces to $\sum_{i=1}^N \mathbf{s}(\mathbf{y}_i)$, computed once during preprocessing, allowing inference to proceed entirely in the ERGM mean value parameter space with negligible additional cost as new networks are added. Using a conjugate prior and Laplace approximation, the method avoids MCMC entirely and is substantially faster than exchange-based approaches. Its main limitation is the equivalent vertices assumption, which excludes inference from a single large observed network.

\subsection{Weighted Networks}

\citet{cai:gol20} extended Bayesian ERGMs to weighted networks by representing a network with $W$ discrete edge weights as a stack of $W-1$ binary layers obtained through thresholding. Each layer is modelled as a binary ERGM generated by dissolving edges from the previous layer, preserving the interpretation of standard ERGM statistics within a hierarchical framework. Inference extends the approximate exchange algorithm \citep{cai:fri11} to the multilayer setting through layer-wise auxiliary network simulation. This provides a highly interpretable Bayesian framework for weighted networks without requiring a dedicated weighted ERGM specification. \citet{fan:whi24}, in a comparative review of weighted ERGM approaches for neuroimaging data, identify the Multi-Layered ERGM of \citet{cai:gol20} as the most practically suitable method due to several criteria including its robustness and interpretability.

\subsection{Longitudinal Networks}

\citet{kos:lom13} and \citet{kos:cai:lom15} developed a two-paper methodological contribution to Bayesian longitudinal ERGMs, extending the framework to continuous-time and temporally evolving networks. The first paper introduces a continuous-time ERGM without equilibrium assumptions, enabling dynamic modelling of tie formation with endogenous dependence and revealing locally hierarchical structures. The second paper extends this to a full Bayesian hierarchical specification, jointly modelling the initial network and allowing structural parameters to evolve over time. Inference is carried out via an extension of the approximate exchange algorithm adapted to the longitudinal setting.

A recent extension of dynamic Bayesian ERGMs to signed networks, in which ties can be either positive or negative to represent supportive or antagonistic relationships, is provided by \citet{cai:gol25a}. Building on a multilayer formulation, the authors develop a separable dynamic ERGM that extends discrete-time separable temporal ERGMs to model the evolution of tie polarity conditional on the underlying interaction process.

\subsection{Prior Specification}

In most Bayesian ERGM implementations, a multivariate Gaussian prior $\pi(\boldsymbol{\vartheta}) = \mathcal{N}(\mathbf{0}, \sigma^2 \mathbf{I})$ is placed on the structural parameters, with $\sigma^2$ chosen to be weakly informative relative to the scale of the sufficient statistics. This choice is computationally convenient, it yields log-concave contributions to the acceptance ratio, and acts as a regulariser that reduces the risk of degeneracy in the posterior. The conjugate prior construction of \citet{yin:but22} offers an alternative in the graph-set setting, providing closed-form posterior updates under exchangeability. Sensitivity to prior hyperparameters is rarely examined systematically in applied work, and represents an under-explored aspect of Bayesian ERGM practice; simulation studies suggest that inference on structural parameters is generally robust to moderate changes in $\sigma^2$, but that model selection criteria such as the ELBO can be more sensitive when competing models differ substantially in dimension.

\subsection{Shrinkage Priors and Functional Extensions}

\citet{par:jeo:shi:jeo:jin22} propose a Bayesian shrinkage framework for functional network models, a class of ERGMs in which the sufficient statistics are replaced by smooth functions of the network, making the model applicable to longitudinal item response data where multiple systems of nodes with different types of local dependence coexist across time. The key challenge is that standard ERGM sufficient statistics are not directly applicable in this bipartite, multimode setting; the functional extension accommodates time-varying network structure by modelling the ERGM parameters as smooth functions of time, with Bayesian shrinkage priors, specifically, spike-and-slab and horseshoe priors, placed on the functional coefficients to perform simultaneous parameter estimation and variable selection across time points. Inference is carried out via an extension of the double Metropolis-Hastings sampler of \citet{lia:jin13}, adapted to the functional parameter space. The method is demonstrated on longitudinal educational assessment data, where it recovers interpretable time-varying item interaction networks whilst automatically shrinking negligible dependencies to zero.

\section{Software}\label{software}

User-friendly software has been crucial to the development of Bayesian ERGM methodology. Most implementations rely on R, with the \texttt{ergm} package \citep{ergm4}, part of the \texttt{statnet} suite \citep{statnet}, providing core model specification and network simulation tools.

The main package for Bayesian ERGM analysis is \texttt{Bergm} \citep{cai:fri14}, which supports Bayesian parameter inference, missing data imputation, model selection, and goodness-of-fit diagnostics. It implements the approximate exchange algorithm \citep{cai:fri11}, adjusted pseudo-likelihood methods \citep{bou:fri:mai17, bou:fri:mai18}, inference under missing data \citep{kos:rob:pat10}, and model evidence estimation. Recent versions have substantially improved efficiency and usability. 
%The \texttt{hergm} package \citep{sch:lun18} provides Bayesian inference for hierarchical ERGMs with block structure, supporting both parametric and nonparametric priors on block membership. 

Beyond R, \texttt{PNet} \citep{wan:rob:pat:kos09} supports Bayesian estimation for several ERGM classes, while variational inference code accompanying \citet{tan:fri20} is available in Julia, although no widely maintained Julia or Python Bayesian ERGM package currently exists.

\section{Discussion and Future Directions}\label{discussion}

Bayesian ERGM approaches offer several important advantages for network analysis, including coherent uncertainty quantification, principled incorporation of prior information, probabilistic model comparison, and a natural framework for hierarchical and complex dependence structures. At the same time, the Bayesian analysis of ERGMs presents considerable computational and modelling challenges, primarily due to the doubly-intractable nature of the likelihood and the increasing complexity of contemporary network data. In response, the field has evolved from early exchange-based MCMC algorithms towards increasingly scalable and theoretically grounded inferential procedures, including variational approximations, surrogate models, and recent simulation-based amortised inference approaches. Modelling  advances have enabled Bayesian ERGMs to address a broader range of relational phenomena, extending beyond static binary networks to temporal, weighted, signed, multi-layer, and population-level settings.

%\subsection{Computational Scalability}

Despite substantial progress, Bayesian inference for ERGMs remains practically confined to networks of at most a few thousand nodes. The principal bottleneck remains the repeated simulation of auxiliary networks, required by exchange-based methods to circumvent the doubly-intractable normalising constant. Since the computational burden increases rapidly with network size and model complexity, no existing Bayesian approach fully resolves scalability limitations. A major emerging direction is Neural Posterior Estimation (NPE), which replaces repeated MCMC sampling with amortised likelihood-free inference. By training normalising flows on parameter--simulation pairs, NPE learns an approximation to the posterior distribution that can subsequently be evaluated via a single forward pass. \citet{fan:whi25a} provide the first systematic implementation for ERGMs, demonstrating substantial computational gains relative to exchange-based algorithms. However, important challenges remain, including poor posterior coverage, sensitivity to the choice of network summaries, boundary effects, and ERGM-specific failure modes such as posterior leakage. Building on this framework, \citet{fan:whi25b} extend NPE to the hierarchical ERGM of \citet{leh:whi24} through Amortised Hierarchical Sequential NPE, enabling inference for large neuroimaging data. Related developments include neural network surrogates that learn the mapping between parameters and sufficient statistics and invert it for fast estimation \citep{mel25}. 

%\subsection{Model Development}

Alongside computational advances, an important frontier concerns the expansion of Bayesian ERGMs to accommodate increasingly rich forms of network dependence and heterogeneity. Much of the early Bayesian literature focused on static binary networks, but contemporary applications frequently involve temporal, weighted, signed, multi-layer, or population-level relational data. Another promising direction concerns hierarchical and population-level ERGMs, which seek to model multiple related networks while borrowing strength across individuals or groups. Hierarchical formulations are particularly relevant in applications involving replicated networks, although computational complexity and model assessment remain challenging. Related open problems include the principled incorporation of latent heterogeneity, uncertainty in network boundaries, and mechanisms for missing or partially observed relational data.

\subsection*{Acknowledgements} We thank the organisers of the workshop \textit{New Trends in Statistical Network Analysis}, Carsten Jentsch, G\"{o}ran Kauermann, and Alexander Kreiss, where this work originated, for fostering a stimulating and productive scientific exchange.

\backmatter

%\bibliography{myref}

\end{document}